# Quantum confinement of exciton-polaritons in a structured (Al,Ga)As microcavity

**Alexander S. Kuznetsov\*, Paul L. J. Helgers, Klaus Biermann and Paulo V. Santos**

Paul-Drude-Institut für Festkörperelektronik,
Leibniz-Institut im Forschungsverbund Berlin e. V.,
Hausvogteiplatz 5-7, 10117 Berlin, Germany

**Abstract**. The realization of quantum functionalities with polaritons in an all-semiconductor platform requires the control of the energy and spatial overlap of the wave functions of single polaritons trapped in potentials with precisely controlled shape and size. In this study, we reach the confinement of microcavity polaritons in traps with an effective potential width down to 1 μm, produced by patterning the active region of the (Al,Ga)As microcavity between two molecular beam epitaxy growth runs. We correlate spectroscopic and structural data to show that the smooth surface relief of the patterned traps translates into a graded confinement-potential characterized by lateral interfaces with a finite lateral-width. We show that the structuring method is suitable for the fabrication of arrays of proximal traps, supporting hybridization between adjacent lattice sites.

## 1. Introduction

The strong coupling of the resonant light field (cavity mode) and quantum-well excitons in a planar semiconductor microcavity (MC) results in a Rabi splitting leading to the creation of exciton-polariton quasi-particles (simply, microcavity polaritons, MPs) [1–4]. MPs display a wide range of attractive properties arising from their dual light-matter nature, such as bright directional emission, lasing [5,6], macroscopic spatial coherence at Kelvin-temperatures [7], Bose-Einstein-like condensation [8] (BEC) and strong nonlinearities [8–10]. MPs offer a pathway to bridge conventional micro-electronics and quantum technologies in an all-semiconductor platform [11]. Many fundamental effects have been demonstrated with polaritons, such as topological edge-states [12], Josephson oscillations [13], superfluidity [14], lattice effects [15–17] etc.

In order to achieve quantum functionalities (e.g., qubits [18], topological circuits [19] and quantum simulators [20]) one needs polariton confinement potentials as well as engineered lattices supporting polariton-interactions and coherent control down to single polariton level. Specifically, the low effective mass of MPs (typically on the order of $10^{-5} - 10^{-4}$ of the free electron mass, $m_e$) and, consequently, their μm-sized thermal de Broglie wavelength (on the order of a few μm at 4 K), favours the observation of confinement effects under μm-size modulation potentials. Different approaches have been reported to create polariton modulation potentials based on the energy control of MPs excitonic or photonic components [21–27]. A particularly attractive approach for the formation of MP traps and lattices is the patterning by shallow etching of the MC spacer followed by overgrowth by molecular beam epitaxy (MBE) [28–31]. Here, the MBE growth is interrupted after the deposition of the MC spacer (cf. Figure 1) – the sample is then structured by shallow etching and reintroduced in the

\*Corresponding author: kuznetsov@pdi-berlin.de

MBE chamber for the growth of the upper distributed Bragg reflector (DBR). This structuring approach is particularly useful since it affects only the MPs photonic modes while keeping intact quantum wells (QWs) embedded into the MC spacer, which are very sensitive to etching-induced defects. In addition, it allows the fabrication of traps and lattices of arbitrary shape and large confinement potentials. Finally, the structuring procedure is compatible with the formation of tuneable lattices and traps via the dynamic modulation by acoustic fields [25].

While previous studies have demonstrated the confinement of MPs in traps as well as inter-trap interactions in lattices produces by MBE overgrowth [28,29,31], little attention was given to the correlation of the physical shape of the traps and the spatial profile of the resulting MP trapping potential. One of the reasons is that most of these studies have addressed traps with dimensions of a few µm, for which the dimensions of the confinement potential are close to the ones defined by the lithographically patterned areas on the MC spacer. The situation changes significantly for smaller traps. In particular, the MBE overgrowth process on the patterned surface is not conformal and also directionally anisotropic [32–34]. As a result, the overgrowth smoothens the lateral interfaces and distorts the shape of the structures defined by etching the MC spacer. The precise knowledge of resulting MP potential profiles is critical for design of tunnel-coupled traps with µm sizes. Like-wise, the spatial extent of the potential imposes a limit on the on-site interaction-strength of single-polaritons, which for unconfined polaritons lies in the range of 0.01 meVµm$^2$ [35,36] to 1 meVµm$^2$ [37].

In the present work, we show the ability to confine polaritons in arrays of µm-sized intra-cavity traps created by patterning an (Al,Ga)As MC in-between growth steps by molecular MBE. We present a detailed study of the impact of the structuring and subsequent MBE overgrowth on the energy of the confined states, aiming to determine the minimum confinement dimensions that can be achieved by the process. The investigations were carried out by combining structural data obtained by atomic force microscopy (AFM) with spectroscopic studies using spatially resolved reflection, and spatially and momentum resolved photoluminescence (PL) to probe the MPs energy levels and wave functions. We show that the MBE overgrowth process gives rise to a graded potential profile for the lateral interfaces of the traps with different interface widths along the [-110] and [-1-10] surface directions of the (001) GaAs substrate. An important result is a quantitative model for the lateral profile of the interfaces, which accurately predicts the shape of the MP confinement and accounts for the energy levels, and explains non-degenerate energy levels of confined polaritons in anisotropic 2D and 3D traps. We demonstrate that the smooth potential shape helps to achieve a spatial overlap of the MPs wavefunctions in arrays of proximal traps. The results presented here are particularly relevant for polariton lattices for quantum simulations, which require small traps sizes (<1µm) for the confinement of single polaritons per lattice site as well as small lattice constant for inter-site interactions.

In the following section (Sec. 2), we describe the design and processes for the fabrication of the structured MC as well as the structural and spectroscopic techniques used in the studies. The experimental results are presented in Sec. 3: here, we start by analysing the impact of the MBE overgrowth on the shape of the mesas (Sec. 3.1) and by addressing the polariton energy levels in extended etched and non-etched areas (Sec. 3.2). Experimental results demonstrating MP confinement in 2D and 3D and their relationship to the confinement dimensions are then presented in Secs. 3.3.1 and 3.3.2, respectively. Section 3.4 presents a direct spectroscopic evidence of coupling between proximal traps in arrays. In Sec. 3.5 we discuss the condensation of MP in a single trap as well as the linewidth of the emission lines. Section 4 summarizes the main results of the studies.

## 2. Experimental Details

The structured (Al,Ga)As MC was grown on a 2-inch GaAs (001) substrate by MBE (cf. Figure 1). We first grew a 4.43 µm-thick composition-graded lower DBR consisting of 36 λ/4 (λ is the optical wavelength) pairs of Al$_{0.15}$Ga$_{0.85}$As/Al$_x$Ga$_{1-x}$As with Al composition *x* continuously reducing from 0.80 in the first stack to 0.45 in the last stack. The first 120 nm of the Al$_{0.30}$Ga$_{0.70}$As MC spacer were then deposited including six 15 nm-thick GaAs QWs placed at the antinode positions of the MC optical mode. The structure was subsequently capped by a 170 nm-wide $Al_{0.15}Ga_{0.85}As$ layer spacer. The sample was taken out of the MBE chamber and then patterned by means of photolithography and wet chemical etching. The latter creates mesas with a nominal height of 12 nm of different shapes in the exposed spacer layer, thus inducing a lateral modulation of the cavity thickness and, therefore, of the cavity energy in the final structure. The etching depth was selected to blue-shift the optical cavity mode in the etched areas by 9 meV (4.5 nm) with respect to the non-etched regions. The upper surface of the etched layer corresponds to a node of the optical cavity mode of the whole structure. In this way, we minimize potential impact of roughness or impurities introduced by the *ex-situ* patterning on optical properties of the structure. Furthermore, the shallow patterned layer is located more than 140 nm above the QWs, so that we can safely assume that they remain unaffected by the processing.

The sample was then reinserted into the MBE system, cleaned by exposure to atomic hydrogen (30 min at 450 °C at a pressure of $2\times10^{-5}$ mbar), and overgrown with a $\lambda/4$ $Al_{0.15}Ga_{0.85}As$ followed by the upper DBR. The latter consists of 20 $\lambda/4$ pairs of $Al_{0.15}Ga_{0.85}As/Al_{0.75}Ga_{0.25}As$. The morphology of the overgrown sample surface was observed by atomic force microscopy (AFM) using a Bruker Dimension Edge system.

The sample was designed to be in the strong coupling regime both in the etched and non-etched regions, leading to MPs in these two regions with different energies and photon/exciton contents. The latter is used to create 2D (wires) and 3D (dots) confinement in non-etched areas surrounded by etched barriers, which were probed by low-temperature reflection and photoluminescence (PL).

Reflection and photoluminescence (PL) spectra of different regions of the sample were recorded at 10 K using a liquid nitrogen-cooled CCD camera (Princeton Instruments) coupled to a spectrometer (SPEX 750M). The spatially or momentum resolved PL studies were performed under excitation by a focused laser beam from either a tuneable Sirah Matisse Ti-Sapphire CW laser or a quasi-CW laser-diode (repetition frequency = 40 MHz) emitting at 635 nm. The 2D images, recorded on the CCD, map the spectral distribution of the reflectivity and PL along the image cross-section defined by the spectrometer slit.

## 3. Results and Discussion
### 3.1. Structural properties

The overgrowth of the structured spacer results in a measurable relief of the sample surface. The latter was probed by scanning an AFM tip over the surface of the patterned areas, as shown for rectangular (i.e., wire-like) surface mesas in Figure 2(a). The nominally 12 nm-thick mesas defined in the MC spacer give rise to a surface relief with a height of 15 nm after the MBE overgrowth. For large mesas (e.g., Figure 2(b)), its dimensions correspond to the ones defined in the spacer region by photo-lithography and the etching process. For smaller mesas, however, the surface structures reproduce the intended shape along the $y\|[-1-10]$ surface direction but their edge profiles become broader along the perpendicular $x\|[-110]$ direction (Figures 2(c) and 2(d)).

In order to quantify the changes in the shape, we first compared the size of the surface mesa with the dimensions of the etched structures. Figure 2(e) shows an exemplary AFM profile along the [-1-10] direction of the nominally $w_{y,nom} = 8$ µm-wide wire of Figure 2(b). From the measurement we extract the actual wire width ($w_y$) along $y\|[-1-10]$, defined as the full-width at half-maximum (FWHM) of the AFM profiles. The dots in Figure 2(f) display the FWHM of the AFM profiles $w_x$ and $w_y$ along [-110] and [-1-10] direction of the surface profiles, respectively, as a function of the nominal widths of the mesa defined in the MC spacer. The measured points of $w_y$ follow closely a linear relation $w_y = w_{y,nom} + \Delta w_y$, as confirmed by the small value for the deviation $\Delta w_y < 0.3$ µm with respect to the line $w_y = w_n$ (diagonal dashed line). The latter indicates that the surface mesa dimensions along $y\|[-1-10]$ are close to the ones defined in the MC spacer. In the case of the $x\|[-110]$ direction the linear slope of $w_x$ is also preserved. Note, however, that $w_x$ is on average 1 µm larger than the nominal width, thus indicating a substantial size increase along this direction.

In addition to the change in average dimensions, a second consequence of the shape anisotropy is a change in the abruptness of the mesa edges. The latter will be quantified in terms of the interface widths $L_x$ and $L_y$ along x and y, as defined in Figure 2(e). The widths of the left ($L_{left}$) and right ($L_{Right}$) lateral interfaces were found to be approximately the same for both crystallographic directions. The value of $L_i$ (i = x, y) was determined as the average value ($L_{left\_i}+L_{Right\_i}$)/2. The dependence of $L_i$ on the nominal width of the wires is summarized in Figure 2(f). The lateral interfaces are much more abrupt along y (with $L_y \leq 1$ μm) than along x (with $L_x \approx 2$ μm). This difference in lateral interface widths is attributed to a combination of three factors. The first is the atomic incorporation rate at the shallowly inclined etched sidewalls (about 10 degrees with respect to the sample plane) during MBE growth, which is distinct from the one for the normal (001) surface. The other two factors are associated with the (2x4) reconstruction of the growing surface, which leads to (i) an enhanced diffusion of adatoms along x||[-110] and, simultaneously, to (ii) a more efficient incorporation at step edges along the perpendicular direction [32–34]. The resulting effect is a smoothening of the lateral interface profile, which is much more pronounced for the x than for the y direction.

According to Figure 2(e), interface smoothening is essentially independent of the width of the mesas. The growth-induced shape anisotropy of the interface has, therefore, a stronger impact on small structures defined within the spacer. In particular, small square mesas, see Figure 3(a), turn into rectangular structures with rounded shape at the sample surface, as illustrated in Figure 3(b) for an etched mesa with a nominal width of 4 μm. The cross-sections of the mesa height displayed in Figure 3(c) and 3(d) show that the edges profiles along the x||[-110] direction are much smoother than the ones along y||[-1-10]. We will show in Sec. 3.3.2 that the shape anisotropy has a pronounced effect on the symmetry and emission energy of the confined MP levels.

## 3.2. Polariton energy levels

Optical reflectivity at 10 K was employed as the basic characterization tool to access the optical properties of the etched and non-etched areas of the sample. Overall, the cavity stopband covers spectral range from 1.49 eV to 1.58 eV. The spatial reflectivity map displayed in Figure 4(a) was recorded by imaging a 110 μm-long and 3 μm-wide area of the sample surface on the slit of the spectrometer. The measured sample area contains spatially extended (width ≈ 40 μm) non-etched (nER) and etched regions (ER) separated by two narrow (≈ 10 μm) wire-like ER and nER regions. Both the nER and ER areas of the map show pronounced resonances close to the centre of the stopband. The reflectivity spectra of extended nER and ER areas, spatially integrated over their respective regions, are shown in Figure 4(b) and (c), respectively. As expected, the resonances in the ER are blue-shifted with respect to the ones in the nER.

Figure 4(d) and 4(e) show momentum-resolved PL spectra recorded at 10 K on extended nER and ER regions, respectively, under non-resonant excitation ($E_{Exc}$ = 1.631 eV) and low excitation density (0.3 W/cm$^2$). The PL maps of nER and ER regions close to the Γ-point (i.e., for in-plane wave vectors $k_{in-plane}$ = 0) are characterized by parabolic-signatures typical of the lower and upper MP branches. The emission of the lowest MP branches dominates PL spectra of both regions. The experimental dispersions are fitted well by solving the eigenvalue-problem for three coupled oscillators representing the bare optical ($C_i$, i = {nER, ER}) mode of the MC as well as the bare heavy-hole ($X_{HH}$) and light-hole ($X_{LH}$) excitonic resonances of the QWs (note that the bare excitonic energies are the same in the nER and ER areas). For the fits, we set the energy for the excitonic resonances to $X_{HH}$ = 1534 meV and $X_{LH}$ = 1540.6 meV and varied the Rabi-splitting ($\Omega_{HH\_i}$, $\Omega_{LH\_i}$) and cavity detuning $\sigma_{C\_i} = C_i - X_{HH}$ in both regions of the sample. The best match between the experiment and calculation was obtained for the parameters listed in Table 1. The Hopfield coefficients obtained from the fits were used to determine the effective masses of the lowest polariton branches listed in the last row of the table.

*Table 1: Parameters for the non-etched (nER) and etched (ER)*

| Parameter | Non-etched region (i=nER) | Etched-region (i=ER) |
|---|---|---|
| $X_{HH}$ | 1534 meV | 1534 meV |
| $X_{LH}$ | 1540.6 meV | 1540.6 meV |
| $\sigma_{Ci} = C_i - X_{HH}$ | (-5.4 ± 0.2) meV | (4.3 ± 0.3) meV |
| $\Omega_{HH\_i}$ | (7.6 ± 0.2) meV | (6 ± 0.2) meV |
| $\Omega_{LH\_i}$ | (5.6 ± 0.2) meV | (4.6 ± 0.2) meV |
| $m_{p\,LP}$ | (5.5 ± 0.3)×10$^{-5}$m$_e$ | (2 ± 0.1)×10$^{-4}$m$_e$ |

Using the results in Table 1 we can better understand the mode-structure of the spectra of nER and ER shown in Figure 4(b) and (c), respectively. In the non-etched regions, the optical cavity mode mainly couples to $X_{HH}$ states – its interaction with the $X_{LH}$ is weak ($C_{nER} - X_{LH} \approx 12$ meV > $\Omega_{LH\_nER}$) but still with sufficient oscillator strength to enable the observation of this mode within the stopband. In the etched regions, the cavity resonance-energy shifts to higher values, thus leading to comparable light-matter coupling between the $X_{HH}$ and $X_{LH}$ states.

The effect of etching on the optical properties is well-reproduced by transfer matrix calculations of the optical reflection using the parameters listed in Table I. Since etching only changes the optical resonance energy $C_i$, the bare excitonic resonances ($X_{HH}$ and $X_{LH}$) remain constant over the sample surface. The same set of parameters was used for nER and ER areas, apart from the spacer thicknesses in nER and ER areas, which determine the bare photon energies $C_i$, $C_{nER} = 1528$ meV and $C_{ER} = 1537.4$ meV. The quality factors of the bare photonic modes of the simulated cavity are $Q_{nER} = 5451$, $Q_{ER} = 4631$. The reflectivity calculations account for the dielectric contribution from the excitonic $X_{HH}$ and $X_{LH}$ resonances of the QWs. For that purpose, the heavy-hole (light hole) absorption cross-section was modelled by a Lorentzian line centred at $X_{HH}$ ($X_{LH}$) with a full-width at half maximum (FWHM) of 1.1 meV (1.2 meV) and amplitude of 9.2×10$^4$ cm$^{-1}$ (3.8×10$^4$ cm$^{-1}$). Figures 4(b) and 4(c) compare measured (thick black lines) and calculated spectra (thin red lines) of the spatially extended non-etched and etched areas, respectively. To improve the fitting, the thicknesses of the lower (upper) DBR layers were slightly reduced (increased) by 0.6 % (1.2 %) with respect to the nominal layer thicknesses to better reproduce the overall shape of the stop band in the reflectivity spectrum. In addition, the difference in the spacer layer thickness of the non-etched and etched areas was set to 14 nm, which is only slightly larger than the intended etching depth of 12 nm.

The numerical simulations reproduce nicely the energies and amplitudes of the measured spectral features of both the etched and non-etched regions. One discrepancy is the larger measured linewidth of the lowest resonance in the non-etched region (Figure 4(c)). The latter is an artefact due to the large numerical aperture of $NA = 0.28$ of the 20$x$ objective used to record the data. Specifically, this NA value corresponds to a collection angle of ±16º with respect to the surface normal (in air), which translates into ±2.15 μm$^{-1}$ in-plane momentum range. According to Figure 4(d), the latter yields approximately 3 meV energy range. In contrast, the numerical transfer matrix simulation was done only for the normal incidence. Thus, in the calculated spectrum, the linewidth is given by the cavity quality factor and excitons parameters.

Finally, a close examination of Figure 4(a) reveals that the polariton states within one area (etched or non-etched) are normally confined to this area. They can, however, tunnel through narrow (10 μm wide) stripes separating two equal (nER or ER) areas.

### 3.3 Confined polariton states
The results in Table I show that there is roughly $\Delta E_{LP} = LP_{ER} - LP_{nER} \approx 6 \pm 0.5$ meV-wide energy gap at the Γ-point between the lower polariton (LP) branches of the non-etched (LP$_{nER}$) region and etched (LP$_{ER}$) region. Traps for the quantum confinement of polaritons can then be engineered by sandwiching the non-etched areas between the etched ones. In the following, we investigate the confinement of the lower polariton states by 2D and 3D potentials.

### 3.3.1 Two-dimensional confinement
Figure 5(a-c) displays PL maps spatially resolved along x||[-110] direction of single nER-wires with nominal widths of 10 μm, 6.4 μm and to 3.2 μm surrounded by etched barriers. The PL was excited by

a focused laser spot (FWHM of 10 μm, $E_{Excitation}$ = 1.952 eV) positioned at the center of the wire. In the figures, the horizontal axes are aligned in parallel with the wire cross-sections along the x||[-110] direction, which correspond to the smooth edge profile after the overgrowth (cf. Figure 2). The images display a series of confined states with increasing energy $E_n$, where $n = 1, 2, ...$ denotes the order of the confined state, located between $LP_{nER}$ and $LP_{ER}$. As expected, the energetic splitting between two successive confined levels increases with decreasing wire width. The color-scaled PL intensity modulation yields the profile of the squared MP wave function amplitude $|\Psi_n|^2(x)$ across the wire width $x$. The number of maxima corresponds to the order $n$ of the confined state. The observation of confined states in wires with widths up to 15 μm is a consequence of the very low effective mass $m_p$ of the photonic-like polariton states. As a consequence, the dimensions required to induce quantum confinement increase by a dimensionless factor $A = \sqrt{m_e/m_p}$ relative to the ones for free electrons with mass $m_e$. For typical effective polariton mass $m_p = 10^{-4} * m_e$, $A$ evaluates to 100.

Figure 6(a) summarizes the dependence of the energies $E_n$ of the confined levels on the nominal width $w$ of the wires. The energy scale for $E_n$ is relative to the peak energy $LP_{ER}$ of the lower polaritons in the etched regions, which defines the height of the confinement barrier (equal to $\Delta E_{LP}$ = 6 meV) for states within the non-etched areas. Predictably, the energetic separation between the discrete levels increases with decreasing confinement dimensions. Contrary to the expectations from a square-like confinement potential shape, which applies for large traps (> 5 μm), the energy levels for small traps are approximately equidistant in energy. The latter indicates that the optical confinement produces a graded (rather than a steep) lateral confinement potential with a characteristic width of 1 μm and 2 μm, for y||[-1-10] and x||[-110] directions, respectively, c.f. Figure 2(f). This behavior is attributed to the smoothening of the interface profiles discussed in Sec. 3.1, which ultimately sets the minimum confinement dimensions for the traps.

We now present a phenomenological model to account for the impact of the shape of the lateral interfaces on the energy and wave function of the confined MP modes. We first consider a step-like modulation of the cavity thickness, which should translate into a square confinement potential for the confined modes with barrier height $\Delta E_{LP}$. We determined these energy levels by numerically solving the Schrödinger equation for a square potential with height $\Delta E_{LP}$ and polariton effective mass $m_p$. The calculations were carried out in Fourier space for a periodic super-cell containing the wire potential surrounded by wide barriers (i.e., with dimensions comparable to the wire width). The periodic potential was expanded in 64 plane waves – we have checked that the energy of the confined states remains approximately the same if 32 or more plane waves are used. The dashed lines superimposed on the measured data in Figure 6(a) show the calculated levels $E_n$ for a square potential with a height $\Delta E_{LP}$ = 5.5 meV and polariton effective mass $m_p = 5.5 \times 10^{-5} m_e$ from Table 1. The solutions given by the dashed lines are valid for integer value of n. The model reproduces reasonably well the energy-spectra for the larger wires, $w_x$ = 10, 8 and 6.4 μm, but fails in the case of narrower wires.

The discrepancies between the predictions from the square potential model and the experiments are attributed to the rounded shape of the lateral potential barriers discussed in Sec. 3.1. In order to account for this effect, we have calculated the energy levels assuming a graded interfacial potential $V_l(i)$, where $i = (x, y)$ is the direction index, between the non-etched and etched regions given by:

$$V_l(i) = -\frac{\Delta E_{LP}}{2}\left[Erfc\left(\frac{x - w_i/2}{\sqrt{2}\delta w_i}\right) + Erfc\left(\frac{x + w_i/2}{\sqrt{2}\delta w_i}\right)\right]. \quad Eq.(1)$$

Here, Erfc(ξ) is the complementary error function, $x$ is the spatial coordinate in the direction given by $i$ and $\delta w_i$ describes the effective width of the lateral interface between non-etched and etched regions. The solid lines in Figure 6(a) display the calculated energy levels for the potential of Eq. (1) assuming $i = x||[-110]$, $w_i = w_x = w_{nom} + \Delta w_x$, $\delta w_i = \delta w_x = 0.75$ μm and $m_p = 5.5 \times 10^{-5}$ $m_e$. Here, $w_{nom}$ is the nominal wire width and $\Delta w_x = 1$ μm is a small correction factor introduced to account for changes in width due to the anisotropic overgrowth (cf. Sec. 3.1). The assumed value for $\delta w_x$ yields a length $L_x =$ 3 μm for the lateral interface, which is close to the one determined for the x∥[-110] direction in Sec. 3.1. With these assumptions, the potential reproduces very accurately the energy of all confined states for all wire widths.

Figure 6(b) displays the squared wave functions $|\Psi_n(x)|^2$ for the first confined states (n = 1, 2, 3) in a nER-wire of a width $w_x = w_{nom} + \Delta w_x = 3.2 + 0.5$ μm. The calculations yield spatial extent of $|\Psi_n(x)|^2$ similar to the ones measured in the Figure 5(c). When compared to a square potential, the smooth interface profile given by Eq(1) reduces (increases) the spatial extent of $|\Psi_n(x)|^2$ for modes with energy $\Delta E_{LP} < E_n < -\Delta E_{LP}/2$ ($-\Delta E_{LP}/2 < E_n < 0$). For narrow wires (i.e., with widths comparable to $2\delta w_x$), the shape of the confinement potential $V_l(x)$ as well as the increased penetration into the barriers lead to spectrum with approximately equidistant energy levels. The confined levels of the narrow wires in Figure 6(a) follow indeed very closely this behavior. Finally, we note that the finite width of the lateral interfaces substantially reduces the depth of confinement potential for wire widths $< 2\delta w_x$.

### 3.3.2. Three-dimensional confinement

Three dimensional confinement of polaritons is achieved by surrounding a small non-etched area by etched regions. Panels (a) and (b) of Figure 7 compare PL maps of a *nominally 3.2 μm-wide* wire oriented along the x direction and a *nominally* 3.2×3.2 μm² square trap, respectively. The most apparent difference between the PL maps for the wire and the square trap resides in the PL linewidth. The levels in the wire exhibit a pronounced tail towards high energy due to the normal (i.e., parabolic) MP dispersion along the wire length, corresponding to the absence of confinement. This tail disappears for the 3D confined states of the square traps, leading to spectrally narrow PL lines. We discuss the linewidth of the trap levels in Sec. 3.5. As far as the energy levels are concerned, the additional confinement along the y∥[-1-10] direction in the square trap results in a slightly increased energy of the fundamental state $E_1$ (by approx. 20%) as well as the energy-separation between the confined levels. This behavior contrasts with the expectations for a perfect (no shape anisotropy) square trap, where the energetic separation between consecutive levels should remain the same while the energy of the lowest confined state $E_1$ should double as compared to a wire with the same width. The discrepancy is attributed to the widening of the lateral interfaces along the x∥[-110] direction due to the anisotropic growth, which also breaks the square symmetry of the structure. As a result, the energy-shifts due to confinement *along x* become much smaller than *along y*.

In order to address the impact of the growth anisotropy on the polariton energies, we have calculated the energy levels and wave functions of a square-trap using the numerical model described in the previous section. For the width of the interfaces along x- and y-direction, we have used $\delta w_x = 0.75$ μm as in the previous case and $\delta w_y = 0.45$ μm. Figure 7(d-h) displays $|\Psi_n(x,y)|^2$ maps for the first five confined levels. The anisotropic shape has a small effect on the fundamental state $E_1$ in Figure 7(d), which is nearly isotropic. The anisotropy, however, lifts the degeneracy of levels $E_2$ and $E_3$, which have lobes along the x and y directions, respectively.

In order to compare the wave function profiles with the spatially resolved PL maps, we assume that the emission from each of the confined levels has a Lorentzian spectral shape with a FWHM of 0.1 meV. Figure 7(c) displays the calculated spectrum of these levels (vertical axis) as a function of the projection of their $|\Psi_n(x,y)|^2$ on the x-direction (horizontal axis). The calculated projection reproduces very well the measured PL map of Figure 7(b). In particular, the calculations describe well the lifting of the degeneracy of levels $E_2$ and $E_3$, which gives rise to the additional levels indicated by the arrows.

An important consequence of the parabolic-like (quasi-equidistant spectrum) confinement potential is the fact that while the lowest energy levels are confined to the physical size of the trap, the upper levels can substantially extend into the barrier, cf. Figure 6(b). This is an important consideration for designing arrays of interacting polariton traps.

## 3.4. Hybridization in lattices of polariton traps

The MBE overgrowth provides a flexible way to fabricate arrays of proximal traps with interacting polaritonic states. We show in this section that an array of traps facilitates hybridized MP wavefunctions. As an example, we consider a square array of traps consisting of 4x5 lattice sites, each containing a square trap with nominal dimensions of 1.6x1.6 µm$^2$. The lattice constant of the array is equal to 4.8 µm. Figure 8(a) shows spectrally and spatially resolved PL image of a line of traps within the array obtained under optical excitation by a laser beam with a rather large diameter of approximately 40 µm and excitation power of 30 µW ($E_{Exc}$ = 1.631 eV). Due to the small lattice period and traps size, the lowest levels of the individual traps hybridize to form states with bonding (s-character with energy $E_{1s}$) and anti-bonding (p-character with energy $E_{2p}$) properties. While in the former the electronic wave function concentrates on the trap sites, the wave function of the p-like states peaks in the region between traps. The character of the trap-levels is further evidenced by the momentum-resolved PL image of the same array, measured under the same experiment conditions, c.f. Figure 8(b). Note that both confined levels, $E_{1s}$ and $E_{2p}$, dispersions lie below the parabolic-like dispersion of the barrier states. As expected, the $E_{1s}$ level has *flat dispersion* with the maximum at Γ-point ($k_x$ = 0 µm$^{-1}$). The dispersion of the $E_{2p}$ level has two symmetrical signatures at $k_x \approx \pm 1$ µm$^{-1}$ confirming its p-character. In addition, the dispersion of the $E_{2p}$ level is not flat, which is an indication of an emerging band-structure. Similar results have been reported for a square array of 2 µm large intra-cavity mesa traps separated by less than 5 µm in Ref. [31] and for arrays fabricated by other methods, such as metal islands on the microcavity surface [38], acoustic waves [17], and deep etching of micropillars 1D [39,40] and 2D arrays [41,42].

## 3.5. Condensation in a single trap

In this section we show that the traps support MPs condensates. Figure 9(a) shows normalized PL spectra of a nominally 4×4 µm$^2$ trap recorded at different excitation densities ($P_{Exc}$). In the linear regime (i.e., at low densities), the PL intensity of the confined levels with energies below the barrier has similar values. At larger densities, $P_{Exc}$ > 12 kW/cm$^2$, the PL is dominated by the emission from the lowest confined level. The dependence of the total PL intensity from the lowest confined level and its linewidth (FWHM) on $P_{Exc}$ is summarized in Figure 9(b). The total PL intensity increases exponentially while the linewidth reduces in a step-like fashion to ~ 50 µeV (which is the resolution limit of our spectrometer) for $P_{Exc}$ > 12 kW/cm$^2$, thus indicating the formation of a polariton condensate. As expected, the condensation is accompanied by the blue-shift of PL levels shown in Figure 9(a). One could argue that the sharp lines for $P_{Exc}$ above 12 kW/cm$^2$ in Figure 9(a) arise from photonic lasing rather than from polariton condensation. Note, however, that the emission energy lies approximately 0.5 meV below the bare cavity resonance in an extended region (dashed horizontal line, see also Figure 4(d)). The latter marks a lower limit for the lasing energy, since it neglects the large confinement-induced shift of the lasing modes due to their small effective mass. This proves that the sharp lines arise from polariton condensation in the strong coupling regime rather than from photonic lasing.

Finally, we briefly discuss the linewidth of the trap levels. In the condensed regime, the linewidth drops to less than 50 µeV. Well *below the condensation threshold*, in contrast, the measured linewidth of the lowest confined state in the square trap is approximately 0.3 meV (cf. Figure 9(b)). This linewidth was found to be independent of temperature from 10 K to 1.6 K as well as of the excitation density down to a few nW/cm$^2$. Therefore, we can rule out effects of phonon-induced broadening and broadening due to the polariton-polariton and polariton-exciton reservoir interactions. Keeping in mind that the QWs remain unaffected by etching and that the size of the trap is only a few µm$^2$, we can probably also exclude the inhomogeneous broadening due to the excitonic-part of MPs. In addition, the linewidth of the trapped MPs may be limited by the photonic-resonance. We suggest that, while the expected quality factor of our cavity is Q ≈ 5500 (Sec. 3.2) in extended nER-regions, the Q may be reduced in confined areas with smooth optical interfaces between nER and ER.

## 4. Conclusions

We have investigated polariton quantum-confinement in a laterally modulated (Al,Ga)As MC fabricated by etching and MBE-overgrowth. The lateral modulation of the cavity spacer thickness translates into a lateral modulation of the optical mode of the cavity while maintaining unaffected the quantum-well resonances. This results in a purely optical potential landscape for exciton-polaritons. A reduction in thickness of the spacer by 12 nm translates into a 9 meV blue-shift of the cavity mode energy. Using optical reflectivity, we observed two types of polaritons appearing in the thinner (etched) and thicker (non-etched) regions of the spacer. In the non-etched regions, the polariton resonances are mostly due to the strong coupling between heavy-hole excitons and the cavity mode, while in the etched region both the heavy- and light-hole excitons couple strongly to the cavity mode.

Spatially resolved photoluminescence spectroscopy gives evidence for the confinement of polaritons within traps formed in non-etched areas surrounded by etched ones. As the width of the non-etched region reduces below approximately 15 μm, discrete confinement levels resulting from the quantization of the low-polariton branch are observed. The 2D confinement in wires and 3D confinement in square-like traps, as well as in arrays of traps was demonstrated. The very small polariton effective mass enables us to directly image the confined wave functions and determine the shape of the confinement potential. The latter changes from an approximately square-like potential for wide traps, widths > 5 μm, to a smooth-shape one, for narrower structures. This effect is attributed to the diffusion of atoms during the MBE overgrowth process, which tends to smooth, during the MBE overgrowth, the shape of the lateral interfaces defined by etching. On one hand, the smoothed interface profile leads to stronger confinement of the lowest excited states within the trap. One the other hand, the finite width (of approximately 1 μm) of the lateral interfaces substantially reduces the depth of confinement potential for wire widths less than 2 μm. Well-defined confined states have been observed for effective potential width down to 1 μm, thus demonstrating that the process can reach small polariton confinement dimensions. In the case of a square-like trap, the anisotropic overgrowth results in an elongated rectangular shaped trap, which lifts the confinement degeneracy along the x||[-110] and y||[-1-10] directions. Finally, we demonstrated the coupling of polariton-states in square arrays of small (1.6×1.6 μm$^2$) proximal (lattice constant below 5 μm) traps. Upon condensation of MPs into BEC in a single trap, the PL linewidth reduces from 0.3 meV to 50 μeV (resolution limited). We argue that the linewidth is limited by the deterioration of the cavity quality factor due to the interfaces broadening due to the overgrowth. Due to its single lattice addressability, single level confinement and polariton-polariton interactions via hybridization, the demonstrated system may be suitable for polariton based quantum simulators. The results obtained here provide a solid basis for the design of μm-sized traps as well as lattices with inter-site polariton interactions.

**Acknowledgement**

A. S. Kuznetsov acknowledges the financial support of Deutsche Forschungsgemeinschaft (DFG) in the framework of the project 359162958 "Coherent acousto-optical interactions in structured polariton microcavities. P. L. J. Helgers acknowledges the funding from the European Union Horizon 2020 research and innovation program under the Marie Sklodowska-Curie grant agreement No 642688.

**Figures**

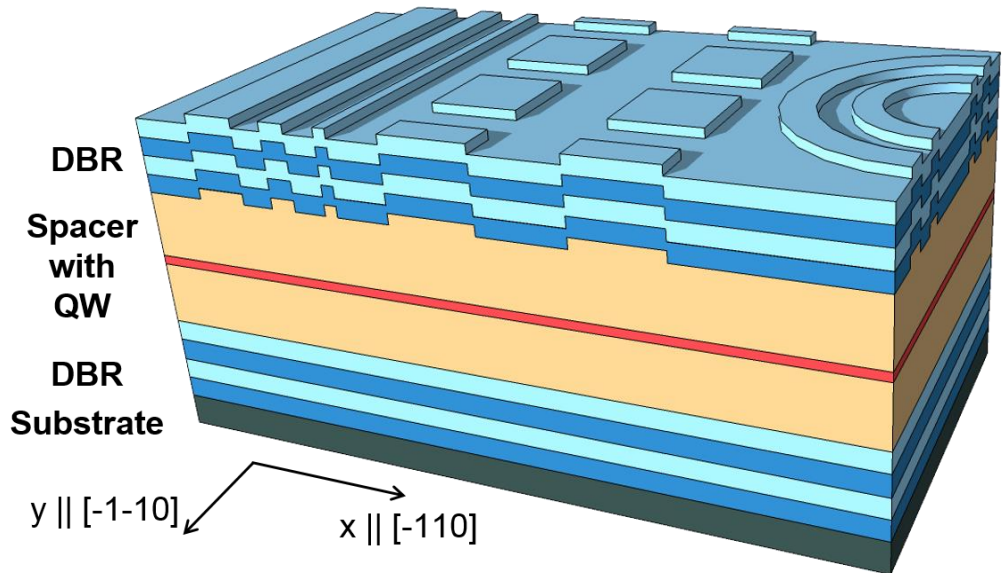

Figure 1. A sketch of the (Al,Ga)As microcavity (MC) with the structured spacer enclosing GaAs QWs grown on a GaAs(001) substrate. The thickness of the spacer in-between the distributed Bragg reflectors (DBRs) was varied by combining etching and overgrowth by molecular beam epitaxy, resulting in regions with different polariton energies.

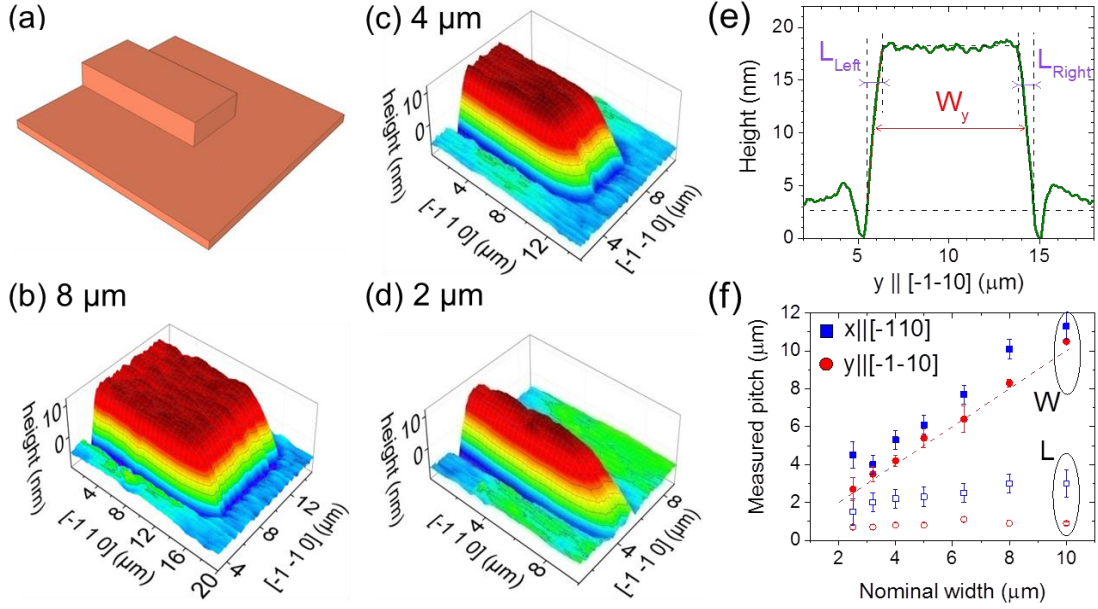

Figure 2. Surface relief of different regions of an overgrown MC. (a) A schematic representation of a wire-like mesa defined in the MC spacer. (b)-(d) Atomic force microscopy (AFM) images of the surface profile of three wire-like mesas defined in the MC spacer with nominal widths of 8 μm, 4 μm, and 2 μm, respectively. The mesa edges expand along the x||[-110] direction due to the anisotropic overgrowth. (e) An exemplary AFM profile of the wire in (b) along y||[-1-10] direction. The $w_y$ and $L_{Left}$&$L_{Right}$ define the full-width half-maximum and interfaces width, $L = (L_{Left} + L_{Right})/2$, respectively. (f) Dependence of the measured values of $w_x$ & $w_y$ and $L_x$ & $L_y$ on nominal wire width. The values were extracted from plots akin in the panel (e).

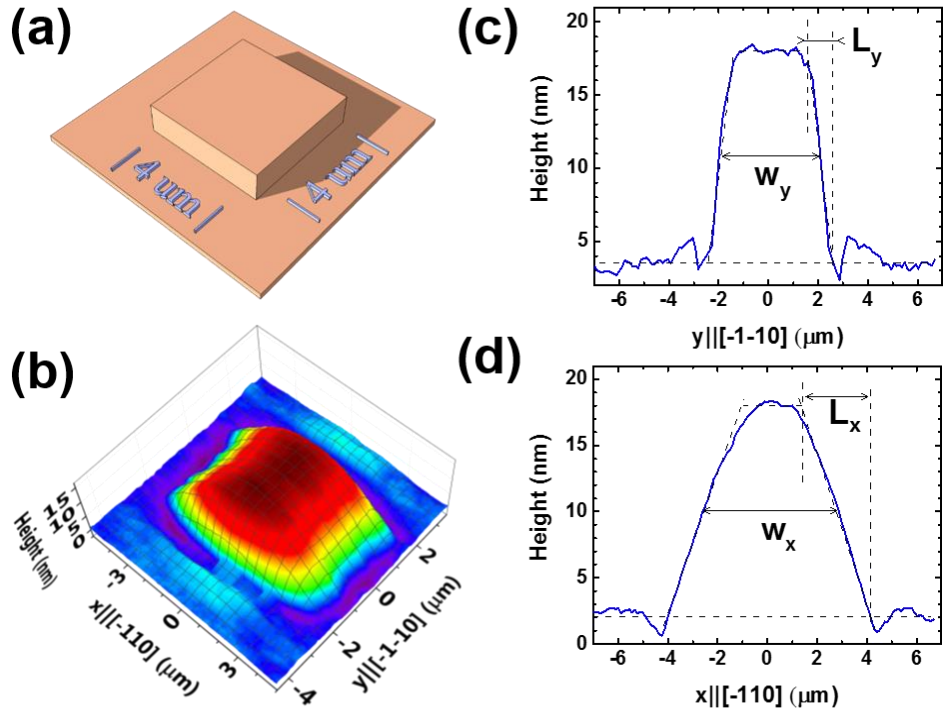

Figure 3. (a) Schematic structure and (b) AFM map of the square-like surface relief arising from the overgrowth of a nominally 4x4 μm$^2$ mesa defined in the cavity spacer. (c) & (d) Height cross-sections of the surface relief in (b) along y∥[-1-10] and x∥[-110], respectively, showing the effects of the anisotropic overgrowth on the mesa shape. The $w_x$ and $w_y$ denote the full width at half maximum along x and y, respectively. $L_x$ and $L_y$ are the corresponding interface widths.

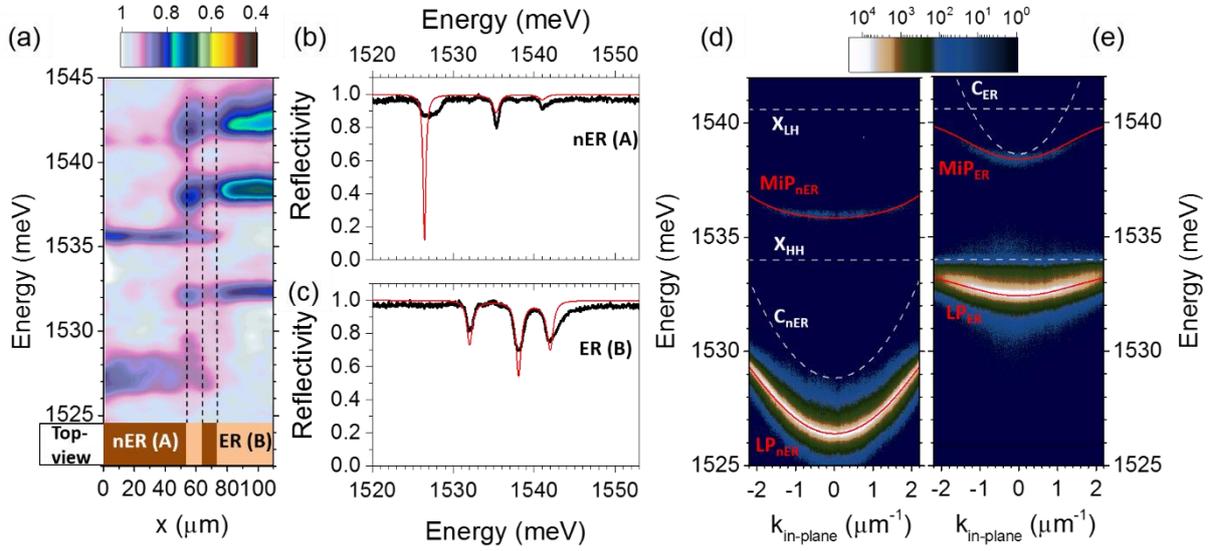

Figure 4. (a) Spatially resolved reflectivity of the structured microcavity at 10K. The horizontal axis corresponds to the orientation of the spectrometer slit along the x∥[-110] direction of the sample surface. The *Top-view* area of the graphic designates the sample surface imaged on the spectrometer slit. (b) & (c) Measured reflectivity spectra (thick black lines) integrated over the nER(A) and ER(B) regions in the *Top-view* part of the panel (a), respectively. Reflectivity spectra of the respective regions calculated by transfer matrix are shown by the thin red curves. (d) & (e) Momentum-space maps of PL of spatially extended nER and ER, respectively. The dashed lines represent bare (uncoupled) dispersion of the heavy-hole ($X_{HH}$) and light-hole ($X_{LH}$) excitons and photonic modes ($C_{nER}$ and $C_{ER}$). Thin red lines correspond to calculated lower branch ($LP_{nER/ER}$) and middle-branch ($MiP_{nER/ER}$) of polariton-dispersions.

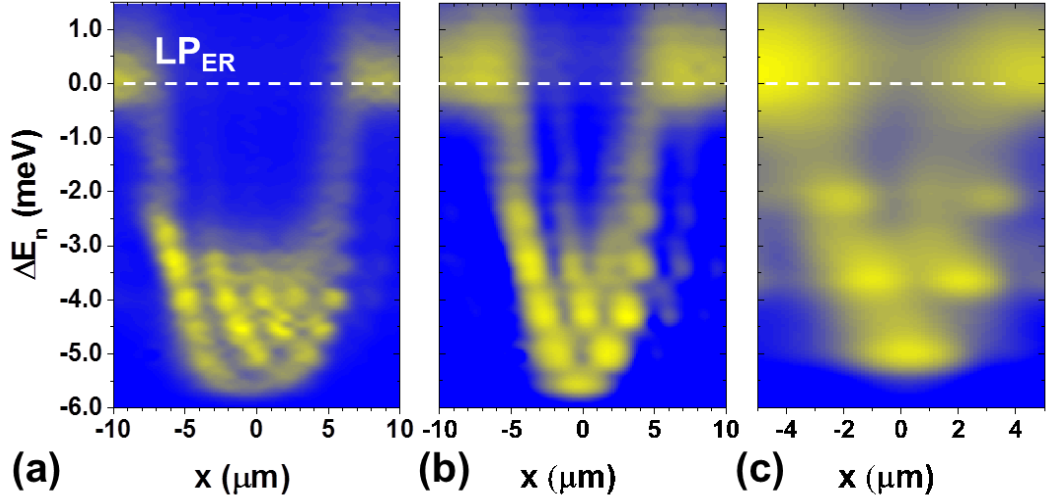

Figure 5: Photoluminescence maps of exciton-polaritons confined in the wire-like nER structures characterized by nominal widths ($w_{nom}$) of (a) 10 μm, (b) 6.4 μm and (c) 3.2 μm, measured along x∥[-110] direction. The energy of the lower polariton state in the etched region $LP_{ER}$ = 1532.5 meV is taken as the reference energy.

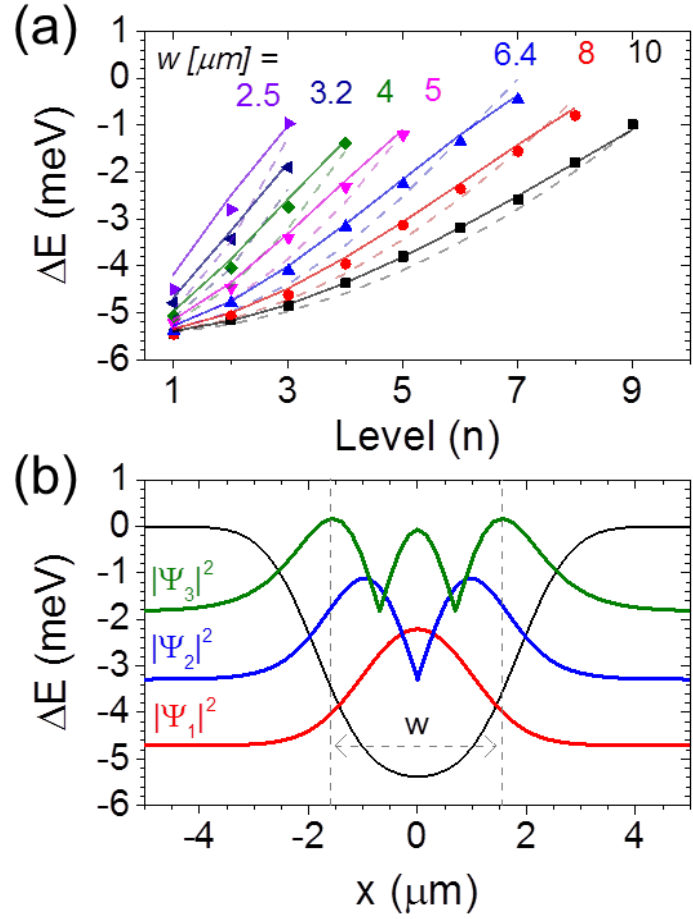

Figure 6. (a) Dependence of the confined energy-levels $\Delta E_n$ on the nominal wire width (symbols) extracted from spatial PL maps akin to the ones in Figure 5, plotted with respect to the barrier energy of 1532.5 meV. The wires are oriented along the $y\|[-1-10]$ direction (i.e., with cross-sections along the fast-growing $x\|[-110]$ direction). The dashed lines were calculated using a square infinitely long wire potential of width ($w_x$) with a barrier height of 5.5 meV and polariton effective mass $m_{eff} = 5.5\times10^{-5}$ $m_e$, where $m_e$ is the free electron mass. The solid lines were calculated using the potential given by Eq. (1). (b) Squared wavefunctions $|\Psi_n(x)|^2$ for the first three confined levels in a wire with nominal width of $w_{nom} = 3.2$ μm calculated using the confinement potential defined by Eq. (1).

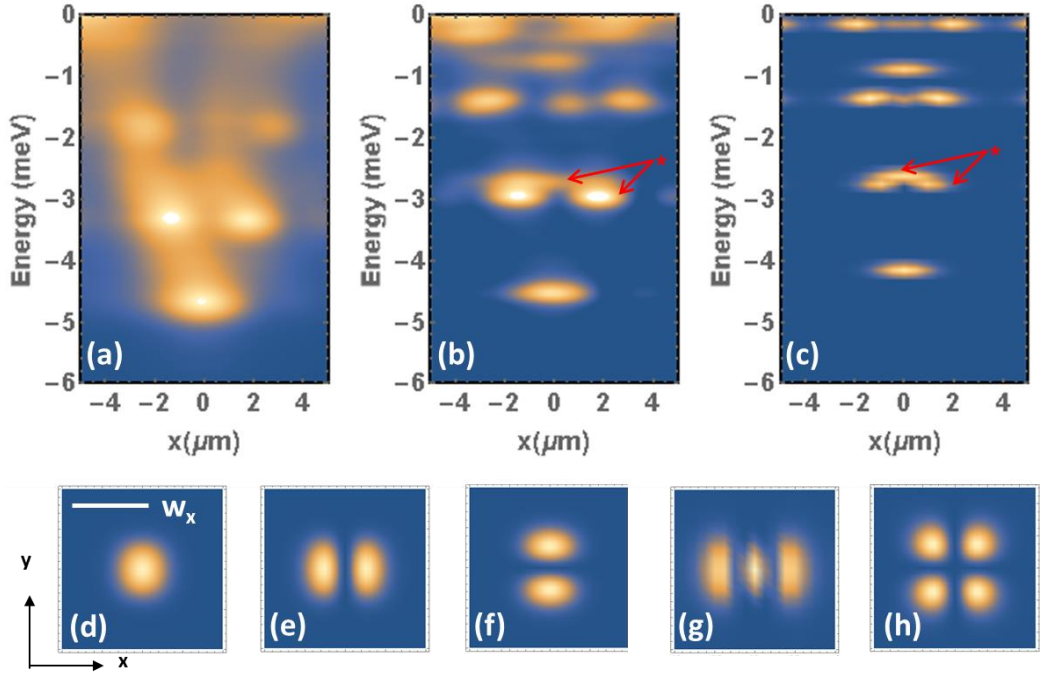

Figure 7. Photoluminescence maps of MPs in non-etched mesas: (a) 3.2 μm wide wire (the same as Figure 5(c)) and (b) nominally 3.2×3.2 μm$^2$ square trap. (c) Calculated projections of the squared wave function $|\Psi_n(x)|^2$ on the x axis for the square trap in (b). Spatial maps of $|\Psi_n(x)|^2$ for the first five confined states in (c) with energies (d) $E_1$ = -4.13 meV, (e) $E_2$ = -2.74 meV, (f) $E_3$ = -2.60 meV, (g) $E_4$ = -1.35 meV, and (h) $E_5$ = -1.31 meV. The red arrows mark two levels (corresponding to the panels (e) and (f)) with degeneracy lifted by the anisotropic shape of the trap. All energies are relative to the barrier energy of 1532.5 meV.

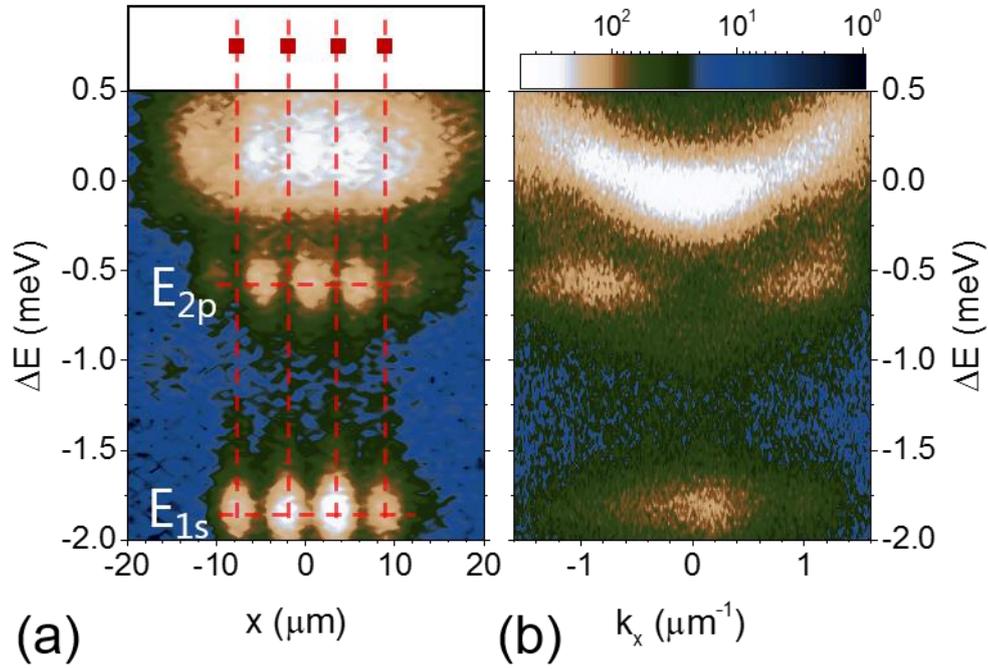

Figure 8. (a) Spatial map of a row of 4 nominally square traps of 1.6×1.6 µm$^2$ size in a 4×5 array with lattice parameter of 4.8 µm, recorded for laser excitation density $P_{Exc}$ = 2 W/cm$^2$, plotted with respect to the barrier energy, $E_{Barrier}$ = 1532.5 meV. The solid squares in the top region of the panel (a) display the spatial arrangement of the traps (squares). The levels $E_{1s}$ and $E_{2p}$ have bonding and anti-bonding symmetries, respectively. (b) $k_x$ cross-section of the array dispersion under the same experiment conditions.

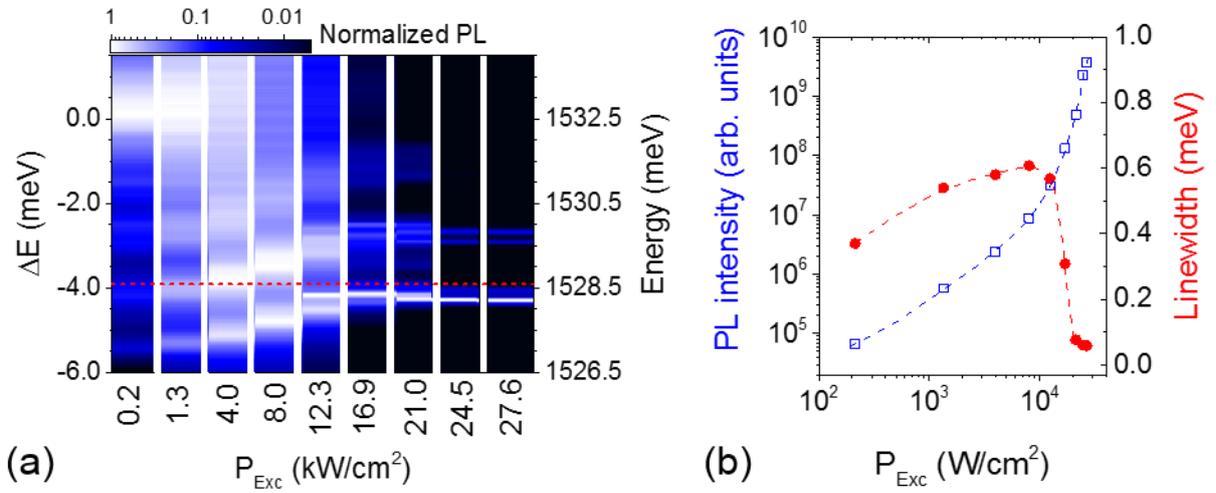

Figure 9. Polariton condensation in a nominally 4×4 μm² trap. (a) Dependence of the trap normalized-PL spectrum (the energy scale is relative to the barrier energy, $E_{Barrier}$ = 1532.5 meV) on the excitation power density ($P_{Exc}$). The PL intensity for each $P_{Exc}$ is normalized to the maximum intensity at that power. The red dashed line gives the energy of the bare cavity mode in the non-etched region (see Table 1). (b) Dependence of the total PL intensity integrated over all levels (open squares) and the linewidth (open circles) of the lowest confined level on $P_{Exc}$. The dashed lines are guides to the eye.